\documentclass[notitlepage,aps,11pt,amsmath,amsfonts,amssymb,nofootinbib,superscriptaddress]{revtex4-1}
\usepackage{enumerate}
\usepackage{graphicx}
\usepackage{mathtools}

\usepackage[colorlinks=true]{hyperref}
\usepackage{amsthm}
\begin{document}

\newcommand{\new}[1]{{\color{blue}\ensuremath{\blacktriangleright}#1\ensuremath{\blacktriangleleft}}}

\title{Curved spacetimes with local $\kappa$-Poincar\'e dispersion relation}

\author{Leonardo Barcaroli}
\email{leonardo.barcaroli@roma1.infn.it}
\affiliation{Dipartimento di Fisica, Universit\`a "La Sapienza''
and Sez. Roma1 INFN, P.le A. Moro 2, 00185 Roma, Italy}

\author{Lukas K. Brunkhorst}
\email{ lukas.brunkhorst@zarm.uni-bremen.de}
\affiliation{Center of Applied Space Technology and Microgravity (ZARM), University of Bremen, Am Fallturm, 28359 Bremen, Germany}
\affiliation{Institute for Theoretical Physics, Universit\"at Hannover, Appelstrasse 2, 30167 Hannover, Germany}

\author{Giulia Gubitosi}
\email{g.gubitosi@imperial.ac.uk}
\affiliation{Theoretical Physics, Blackett Laboratory, Imperial College, London SW7 2AZ, United Kingdom.}

\author{Niccol\'o Loret}
\email{niccolo.loret@roma1.infn.it}
\affiliation{Dipartimento di Fisica, Universit\`a "La Sapienza''
and Sez. Roma1 INFN, P.le A. Moro 2, 00185 Roma, Italy}

\author{Christian Pfeifer}
\email{christian.pfeifer@ut.ee}
\affiliation{Laboratory of Theoretical Physics, Institute of Physics, University of Tartu, W. Ostwaldi 1, 50411 Tartu, Estonia}
\affiliation{Institute for Theoretical Physics, Universit\"at Hannover, Appelstrasse 2, 30167 Hannover, Germany}
\affiliation{Center of Applied Space Technology and Microgravity (ZARM), University of Bremen, Am Fallturm, 28359 Bremen, Germany}

\begin{abstract}

We use our previously developed identification of dispersion relations with Hamilton functions on phase space to locally implement the $\kappa$-Poincar\'e dispersion relation in the momentum spaces at each point of a generic curved spacetime. We use this general construction to build the most general Hamiltonian compatible with spherical symmetry and the Plank-scale-deformed one such that in the local frame it reproduces the  $\kappa$-Poincar\'e dispersion relation. 
Specializing to Planck-scale-deformed Schwarzschild geometry, we find that the photon sphere around a black hole becomes a thick shell since photons of different energy will orbit the black hole on circular orbits at different altitudes.  We also compute the redshift of a photon between different observers at rest, finding that there is a Planck-scale  correction to the usual redshift only if the observers detecting the photon have different masses.

\end{abstract}

\maketitle

\newpage

%\tableofcontents

%%%%%%%%%%%%%%%%%%%%%%%%%%%%%%%%%%%%%%%%%%%%%%%%%%%%%%%
\section{Introduction}
The $\kappa$-Poincar\'e algebra of symmetries, a quantum deformation of the Poincar\'e algebra~\cite{Lukierski:1991pn,Lukierski:1992dt,Lukierski:1993wxa}, is one of the most intensively studied phenomenological models relevant for quantum gravity research. This is mostly because it provides a mathematically consistent example of a relativistic theory with two invariants (the speed of light and the Planck length or energy) and it produces potentially observable effects, such as an energy-dependent propagation velocity of massless particles which may be measured in the observation of $\gamma$-ray bursts at cosmological distances (see~\cite{AmelinoCamelia:2008qg} and references therein). Geometrically, the motion of a particle admitting $\kappa$-Poincar\'e symmetry can be interpreted as happening on a flat spacetime manifold with a curved momentum space enjoying de Sitter symmetry, the Planck scale being related to the curvature of the momentum space itself~\cite{KowalskiGlikman:2002ft, Gubitosi:2013rna}.

As already discussed in \cite{Barcaroli:2016yrl}, in order to make the $\kappa$-Poincar\'e model more suited to describe quantum gravity effects in the cosmological framework, it is necessary to implement the $\kappa$-Poincar\'e dispersion relation on generally curved spacetimes. This entails building a model of intertwined spacetime and momentum space such that in a local frame one recovers the flat spacetime $\kappa$-Poincar\'e dispersion relation. In the local frame the $\kappa$-Lorentz symmetries hold, i.e. the $\kappa$-Poincar\'e symmetries except translations.
By now several steps towards this goal have been achieved. The so called $q$-de Sitter dispersion relation implements the $\kappa$-Poincar\'e dispersion relation on de Sitter spacetime geometry \cite{Barcaroli:2015eqe} and is associated to a quantum deformation of the de Sitter algebra of spacetime symmetries. A first approach to a homogeneous and isotropic spacetime with $\kappa$-Poincar\'e dispersion relation was presented in \cite{Rosati:2015pga} by gluing together slices of its de Sitter spacetime realization. 
Recently we could go even further. In \cite{Barcaroli:2015xda} we interpreted dispersion relations as level sets of Hamilton functions on the cotangent bundle of a spacetime manifold and developed a precise notion of symmetries of dispersion relations. This enabled us to construct the most general homogeneous and isotropic dispersion relation and to identify what we called the qFLRW dispersion relation~\cite{Barcaroli:2016yrl}. It is constructed such that in a local frame the dispersion relation reduces to the $\kappa$-Poincar\'e one, and, when the Planck-scale deformation vanishes it describes the motion of a relativistic particle on Friedmann-Lema\^itre-Robertson-Walker (FLRW)  spacetime. 

Here we show that in fact the $\kappa$-Poincar\'e dispersion relation (and the associated $\kappa$-Lorentz symmetries) can be realized locally on a general curved spacetime. 
Specifically, in section \ref{sec:kappa} we construct a Planck-scale-modified Hamiltonian on a general curved spacetime, such that at every point of spacetime there exists a basis of the cotangent space such that the covariantly-defined dispersion relation takes the standard $\kappa$-Poincar\'e form. 
This characterization is similar to the fact that on every Lorentzian manifold there exist frames of the Lorentzian spacetime metric. In these frames the dispersion relation of point particles on curved spacetime takes the same form as on flat Minkowski spacetime. In \ref{ssec:kappamotion} we use Hamilton equations to work out the motion in phase space of a particle with such Planck-scale-deformed dispersion relation on generic curved spacetime and in \ref{ssec:Redshift} we derive a general formula for the redshift between two observers.
In section \ref{sec:spsym} we specialize our model to the case of a spherically symmetric spacetime, presenting the most general Planck-scale-deformed dispersion relation compatible with these symmetries which reduces to the $\kappa$-Poincar\'e one in the local frame. It contains four free functions depending on the time and radial coordinate satisfying one algebraic constraint. Two of these functions are fixed by the undeformed metric spacetime geometry, as we show in \ref{ssec:kSchw}, where we deal with the Schwarzschild case. In \ref{ssec:kSchwMotion} we work out the Hamilton equations for the Schwarzschild case and finally in  \ref{sub:observable} we compute some observable effects. In particular, we show that the radius of the circular photon orbits around a black hole depend on the photon's energy and that the observed redshift of a photon moving radially in the Schwarzschild geometry is modified with respect to the standard case only if the two observers detecting the photon have different masses.

During this article we use the following notational conventions: Indices $a,b,c,...$ and $\mu,\nu,..$run from $0$ to $3$. Latin indices denote tensor components in manifold induced coordinates, greek indices denote frame induced coordinates $(x,p) \sim P=p_adx^a \in T^*_xM$ of the cotangent bundle. Tensorial objects on spacetime, like a spacetime metric $g$ or a vector field $Z$ are often interpreted as function on the cotangent bundle $g^{-1}(p,p)$ or $Z(p)$ which are defined as these tensors action on the $1$-form~$P$:
\begin{align*}
	g^{-1}(p,p) = g^{ab}p_ap_b,\ Z(p) = Z^ap_a\,.
\end{align*}
The signature convention for the spacetime metric we use is $(-,+,+,+)$. The manifold induced coordinates on the cotangent bundle  satisfy the canonical Poisson relations $\{x^a,p_b\} = \delta^a_b$, all other vanish.

%%%%%%%%%%%%%%%%%%%%%%%%%%%%%%%%%%%%%%%%%%%%%%%%%%%%%%%
\section{$\kappa$-Poincar\'e momentum spaces on curved spacetimes}\label{sec:kappa}
The general framework of Hamilton geometry applied to Planck-scale-modified dispersion relations is discussed in detail in our previous publications \cite{,Barcaroli:2015xda,Barcaroli:2016yrl}. Here we only introduce the basic notions, recalling the connection between dispersion relations and level sets of Hamilton functions on the cotangent bundle of a spacetime manifold. Subsequently, we write down the Hamilton function which implements the $\kappa$ -Poincar\'e dispersion relation for free particles at every point of a generic spacetime and introduce the notion of $\kappa$-Lorentzian symmetry. We then briefly discuss the equations of motion induced by such Hamiltonian and we compute the redshift between any two observers.

%%%%%%%%%%%%%%%%%%%%%%%%%%%%%%%%%%%%%%%%%%%%%%%%%%%%%%%
\subsection{Dispersion relations as level sets of Hamilton functions}
In general relativity local Lorentz invariance is encoded in terms of symmetry transformations on the tangent, respectively cotangent spaces of spacetime. In particular this symmetry manifests itself in the local invariance of the dispersion relation of fundamental point particles, which is given by 
\begin{align}
	 g^{ab}(x)p_ap_b = -m^2\,, 
\end{align}
where $m$ is the invariant mass parameter. On each point $x$ on  spacetime, this dispersion relation is invariant under Lorentz transformations of the momenta $p$ in the following sense: There exist frames $A$ of the metric $g$ such that
\begin{align}\label{eq:Hg}
	H_{g}(x,p) \equiv g^{ab}(x)p_ap_b = \eta^{\mu\nu} A^a{}_\mu(x) A^b{}_\nu(x) p_a p_b = \eta^{\mu\nu}\mathfrak{p}_\mu \mathfrak{p}_\nu \equiv H_\eta(x,\mathfrak{p})\,.
\end{align}
Lorentz invariance manifests itself in the fact that the frame matrix $A$ is not unique. In fact every transformation $\hat A$ which is constructed from  $A$  via the application of a  Lorentz transformation does not change the value of the function $H_{g}(x,p)$.

In previous work \cite{Barcaroli:2015xda,Barcaroli:2016yrl} we have demonstrated that one can interpret the dispersion relation of point particles as level sets of a Hamilton function $H(x,p)$ on the cotangent bundle of spacetime. Then the geometry of the particle's phase space, i.e. the intertwined geometry of spacetime and the point particle's momentum space, can be derived from the Hamilton function. Here we schematically recall the most important features of the Hamiltonian construction of the phase space geometry (further details and explicit examples can be found in \cite{Barcaroli:2015xda,Barcaroli:2016yrl}):
\begin{itemize}
	\item The Hamilton equations of motion are the autoparallel equations of the unique torsion-free Cartan non-linear connection. For non-homogeneous Hamiltonians they include a force-like source term.
	\item The Cartan non-linear connection, uniquely derived from the Hamiltonian, splits the tangent spaces of phase space covariantly in directions along spacetime and along momentum space.
	\item Canonical linear connections, uniquely determined by the Hamiltonian and the non-linear connection, define the  curvature of spacetime and of momentum space, both of which in general depend on both spacetime coordinates and momenta.
	\item For the Hamiltonian $H_{g}(x,p)= g^{ab}(x)p_ap_b$ one obtains the usual Lorentzian metric geometry of spacetime with a flat momentum space.  
\end{itemize}

A dispersion relation thus gives us access to observable predictions by determining the motion of point particles obeying the dispersion relation through the Hamilton equations of motion, encodes the local phase space symmetry in terms of its local invariances and determines the geometry of phase space whose autoparallels coincide with the Hamilton equations of motion.
As mentioned in the introduction, in this article we aim for observable predictions from a modification of the local Lorentz invariant point particle Hamilton function in general relativity. Specifically, we will construct a covariant Hamilton function on a generic curved spacetime whose local symmetry transformations are generated by the $\kappa$-Poincar\'e algebra~\cite{Majid:1994cy}.

%%%%%%%%%%%%%%%%%%%%%%%%%%%%%%%%%%%%%%%%%%%%%%%%%%%%%%%
\subsection{The locally $\kappa$-Poincar\'e Hamiltonian}\label{ssec:genkappa}
The $\kappa$-Poincar\'e  dispersion relation \cite{Majid:1994cy} can be represented as the level sets of the Hamilton function 
\begin{align}\label{eq:kappaflat}
H_\kappa(x,p) = -\frac{4}{\ell^2}\sinh\bigg(\frac{\ell}{2}p_t\bigg)^2 + e^{\ell p_t} \vec p^2\,,
\end{align}
where $p_{t}$ and $\vec p$ are, respectively, the particle's energy and spatial momentum and $\ell=\kappa^{-1}$ is the deformation parameter, such that for $\ell = 0$ the Hamiltonian reduces to the familiar expression of special relativity
\begin{align}\label{eq:minkowski}
H(x,p) = - p_t^2 + \vec{p}^2  = \eta^{ab}p_ap_b\,.
\end{align}
The $\kappa$-Poincar\'e Hamiltonian \eqref{eq:kappaflat} is the $\kappa$-deformation of the flat Minkowski spacetime Hamiltonian~\eqref{eq:minkowski}. The idea is that the Hamiltonian \eqref{eq:kappaflat} determines the effective motion of point particles in a semiclassical regime of quantum gravity, as discussed in \cite{Barcaroli:2015xda}. Hamiltons equations of motion of both imply that all particles move force-free on straight lines in one and the same coordinate system, so both yield particle motion on flat spacetime. The difference is that if one derives the momentum space curvature of \eqref{eq:kappaflat} and \eqref{eq:minkowski} according to the framework of Hamilton geometry outline in the previous section, one finds a non-trivial momentum space curvature in the $\kappa$-Minkowski case, but a vanishing momentum space curvature for Minkowski spacetime.

As mentioned in the introduction, the phenomenological implications of this $\kappa$-deformed Hamiltonian have been widely studied in the literature, however most of the observable effects would be mostly apparent in a cosmological setting or in general in regimes where spacetime curvature can not be neglected. This provides motivation to look for ways to implement the $\kappa$-deformed Hamiltonian on an arbitrarily curved spacetime $(M,g)$, constructing a Hamilton function which locally takes the form \eqref{eq:kappaflat}, so that it is locally $\kappa$-Poincar\'e invariant in the same sense as general relativity is locally Lorentz invariant. In \cite{Barcaroli:2016yrl} we focussed on homogeneous and isotropic FLRW spacetimes. Here we  study the general case. 

Let $(M,g)$ be a globally hyperbolic Lorentzian spacetime and let $Z$ be a normalized globally-defined timelike vector field on $(M,g)$, which can be interpreted as function $Z(p)$ on the cotangent bundle of spacetime $T^*M$,
\begin{align}
g(Z,Z) \equiv g_{ab}(x)Z^a(x)Z^b(x) = -1,\quad Z(p) = Z^a(x)p_a\,.
\end{align}
The $\kappa$-Poincar\'e  deformation of $(M,g)$ is defined by changing the Hamiltonian $H_g$ to $H_{Zg}$ defined by
\begin{align}\label{eq:kappacov}
H_{Zg}(x,p) \equiv -\frac{4}{\ell^2}\sinh\bigg(\frac{\ell}{2}Z(p)\bigg)^2 + e^{\ell Z(p)} (g^{-1}(p,p) + Z(p)^2)\,.
\end{align}
We label the deformed Hamiltonian by the vector field $Z$ since in general different choices of $Z$ lead to different $\kappa$-deformed Hamiltonians. In section \ref{sec:spsym}, where we discuss the spherically symmetric $\kappa$-Poincar\'e phase space, we will see this freedom explicitly. This Hamiltonian can be considered as $\kappa$-deformation of a local Lorentz invariant spacetime to a local $\kappa$-Lorentz invariant one and, in addition, also be derived from a modified theory of electrodynamics as we discuss in Appendix~\ref{app:kappaelectro}.

Performing a power-series expansion in $\ell$ we find
\begin{align}\label{eq:perturb}
H_{Zg}(x,p) = g^{-1}(p,p) + \ell Z(p)(g^{-1}(p,p) + Z(p)^2) +\mathcal{O}(\ell^2)\,.
\end{align}
Thus the zeroth order of $H_{Zg}$ is identical to the Hamilton function which determines the particle motion and the geometry of spacetime in general relativity.

It can be shown that the Hamiltonian \eqref{eq:kappacov}  is locally $\kappa$-Poincar\'e invariant via the following argument. Since $Z$ is a unit-timelike vector, there exists a frame $A$ of the metric $g$ such that $A^a_0\partial_a = Z$, thus $Z(p) = A^a_0 p_a = \mathfrak{p}_0$. Since $A$ is a frame, we can express the metric square of the momenta in this frame as
\begin{align}
	g^{ab}p_ap_b = \eta^{\mu\nu}\mathfrak{p}_\mu\mathfrak{p}_\nu = - \mathfrak{p}_0^2 + \vec{\mathfrak{p}}^2.
\end{align}
Thus with respect to this frame the $\kappa$-Poincar\'e (bicrossproduct basis) Hamiltonian we constructed becomes
\begin{align}\label{eq:kappaframe}
H_{Zg}(x,p) = -\frac{4}{\ell^2}\sinh\bigg(\frac{\ell}{2}\mathfrak{p}_0\bigg)^2 + e^{\ell p_t} \vec {\mathfrak{p}}^2 =  H_{Z\eta}(x,\mathfrak{p}(x)) \,,
\end{align}
which is invariant under the transformations generated by the $\kappa$-Poincar\'e algebra.

The frame $A$ induces a \emph{local and linear transformation} on the momenta such that locally, at every $x\in M$ the Hamiltonian \eqref{eq:kappacov} takes the form \eqref{eq:kappaflat}.  Again, as in the metric general-relativistic case, this transformation is not unique. It can be combined with a $\kappa$-Poincar\'e transformation and the value of $H_{Zg}$ will not change. Thus we conclude that $H_{Zg}$ is locally $\kappa$-Poincar\'e invariant in the same sense as the metric Hamiltonian $H_g$ is local Lorentz invariant. To be precise observe that this local invariance on curved spacetimes excludes the translations from the full $\kappa$-Poincar\'e algebra as transformations. To have a nomenclature for the allowed transformations available we call the remaining elements of the algebra, i.e. the $\kappa$-Poincar\'e boosts and rotations, the $\kappa$-Lorentz algebra.

The intertwined geometry of the smooth point particle phase space, i.e. its linear connections and curvatures, can now be derived according to the framework developed in \cite{Barcaroli:2015xda}. This derivation is beyond the scope of this article which aims for phenomenological implications of the modified dispersion relation induced by the Hamiltonian~\eqref{eq:kappacov}. We also do not study in detail the consequences on  this geometric picture of general non-linear momentum transformations. We only focus on the non-linear momentum transformations that are symmetries of the model, i.e.  that  leave the Hamiltonian invariant, and these are the transformations generated by the $\kappa$-Lorentz algebra.

%Before we continue with the analysis of point particle motion defined by  a locally $\kappa$-Poincar\'e Hamiltonian we like to remark the following: The same line of argument applied above can be generalized to different basis of $\kappa$-Poincar\'{e} Hopf algebra which can be obtained by a nonlinear change of four-momentum coordinates. Those different bases describe different physical models, having a different generator algebra, with a different (first) Casimir operator and thus a different Hamiltonian. The nonlinear change of four-momentum coordinates in general does not affect only  the single particle dynamics, but also the generators' coproducts which formalize particles interactions and composition law of momenta. In particular adopting the right map one can obtain $\kappa$-deformations with a classical Poincar\'{e} algebra sector and undeformed Hamiltonian, with a nontrivial coalgebric sector ({\it id est} $\kappa$-deformed coproducts, antipode and counity), which formalize a physical model with undeformed single particle dynamics and nontrivial interaction vertices. Since in this paper we would like to focus on the dynamics of a single particle in a Schwarzschild geometry with deformed local symmetries, we specialize our investigation on the well-studied $\kappa$-Poincar\'{e} bicrossproduct basis, knowing that any other momentum-basis with a different Hamiltonian would accordingly imply different $g(p,p),\,Z(p)\ \text{and } A$.

Before we continue with our analysis we would like to make a remark concerning different bases of the $\kappa$-Poincar\'e algebra. The same line of argument applied above can be generalized to different bases of the $\kappa$-Poincar\'{e} Hopf algebra, which can be obtained by a nonlinear redefinition of the translation generators. Different bases have in general different Casimir operators and thus they are associated to different Hamiltonians. In particular, there exists a basis with a classical Poincar\'{e} algebra sector and undeformed Hamiltonian. However such a basis is characterized by nontrivial coproducts of the translation generators, that imply nontrivial composition law of momenta in particles' interactions \cite{Gubitosi:2013rna}. Thus such a basis formalizes a physical model with undeformed single particle dynamics and nontrivial interaction vertices. Since in this paper we focus our phenomenological analysis on the motion of a free single-particle in a Schwarzschild geometry with deformed local symmetries, it makes sense to specialize our investigation on the $\kappa$-Poincar\'{e} bicrossproduct basis.

%%%%%%%%%%%%%%%%%%%%%%%%%%%%%%%%%%%%%%%%%%%%%%%%%%%%%%%
\subsection{Particle motion}\label{ssec:kappamotion}
Having implemented the $\kappa$-Poincar\'e dispersion relation locally on a general curved spacetime as level sets of the Hamilton function \eqref{eq:kappacov}, we study the particle motion in phase space which is determined by the Hamilton equations of motion derived from \eqref{eq:kappacov}. 
These are eight first-order ordinary differential equations which are equivalent to four second order ordinary differential equations, the Euler-Lagrange equations of the Lagrangian corresponding to the Hamiltonian in consideration. The transformation of the Hamiltonian representation of the theory to its Lagrangian counterpart is the starting point for finding a Finsler geometric formulation of the $\kappa$-deformed geometry of spacetime, which is investigated in several articles \cite{Amelino-Camelia:2014rga,Lobo:2016xzq,Letizia:2016lew}.

The Hamilton equations of motion of the general $\kappa$-deformed Hamiltonian imply the following relation between velocities and momenta:
\begin{align}\label{eq:dotx}
	\dot x^a 
	&= \bar{\partial}^a H_{Zg}\\
	&= Z^a \bigg[-\frac{2}{\ell} \sinh\big(\ell Z(p)\big) + \ell e^{\ell Z(p)} (g^{-1}(p,p) + Z(p)^2) + 2 e^{\ell Z(p)}  Z(p)\bigg] +2 \, e^{\ell Z(p)}  g^{ab}p_b \nonumber \,,
\end{align}
while the evolution of momenta is given by
\begin{align}\label{eq:deformedgeod}
	\dot p_a 
	&= - \partial_a H_{Zg}\\
	&= p_q \partial_aZ^q \bigg[\frac{2}{\ell} \sinh\big(\ell Z(p)\big) - \ell e^{\ell Z(p)} (g^{-1}(p,p) + Z(p)^2) - 2 e^{\ell Z(p)}  Z(p)\bigg] -  e^{\ell Z(p)}  p_b p_c \partial_a g^{bc} \nonumber\,.
\end{align}			
The latter can be written in an explicitly covariant form with respect to manifold induced coordinate transformation by introducing the Levi-Civita connection of the Lorentzian metric $g$:
\begin{align}  
	\dot p_a	  
	&= p_q \nabla_a Z^q \bigg[\frac{2}{\ell} \sinh\big(\ell Z(p)\big) - \ell e^{\ell Z(p)} (g^{-1}(p,p) + Z(p)^2) - 2 e^{\ell Z(p)}  Z(p)\bigg] + 2 e^{\ell Z(p)} p_c p^b \Gamma^c{}_{ba} \nonumber \\
	&+ p_q \Gamma^q{}_{ab}Z^b \bigg[\frac{2}{\ell} \sinh\big(\ell Z(p)\big) - \ell e^{\ell Z(p)} (g^{-1}(p,p) + Z(p)^2) - 2 e^{\ell Z(p)}  Z(p)\bigg] \nonumber\,.
\end{align}
Observe that this spacetime metric is used here as an available mathematical tool to check the covariance of the equations of motion explicitly. It is not what fundamentally determines the geometry of spacetime, momentum space nor the motion of particles (in fact we can not really separate spacetime and momentum space within the phase space). The fundamental ingredient is the Hamilton function itself and when the Planck-scale corrections are introduced spacetime and momentum space are intertwined so that it is not possible to talk about a spacetime metric on its own.

Reshuffling the terms in the above equations we find
\begin{align}
	&p_q \nabla_a Z^q \bigg[\frac{2}{\ell} \sinh\bigg(\ell Z(p)\bigg) - \ell e^{\ell Z(p)} (g^{-1}(p,p) + Z(p)^2) - 2 e^{\ell Z(p)}  Z(p)\bigg] =\\
	&\dot p_a - 2 e^{\ell Z(p)} p_c p^b \Gamma^c{}_{ba} - p_q \Gamma^q{}_{ab}Z^b \bigg[\frac{2}{\ell} \sinh\bigg(\ell Z(p)\bigg)- \ell e^{\ell Z(p)} (g^{-1}(p,p) + Z(p)^2) - 2 e^{\ell Z(p)}  Z(p)\bigg]\nonumber\,.
\end{align}
Since the Hamilton equations of motion are covariant, i.e. behave tensorial under manifold induced coordinate changes, and since the left hand side of these equations are covariant as well, the right hand side must be covariant. For $\ell\to0$ we obtain, as expected, the geodesic equation in its Hamilton formulation and the usual relation between momenta and velocities in general relativity
\begin{align}
	\dot x^a = 2 g^{ab}p_b,\quad \dot p_a - 2 p_c p^b \Gamma^c{}_{ba} = 0 \Rightarrow \ddot x^a + \Gamma^{a}{}_{bc}\dot x^b \dot x ^c = \nabla_{\dot x}\dot x  = 0\,.
\end{align}
We do not transform the Hamilton equations of motion of the $\kappa$-deformed Hamiltonian into their Euler-Lagrange form explicitly since this is a lengthy calculation not needed for the scope of this article. If needed they can be directly calculated from the Lagrangian corresponding to the $\kappa$-deformed Hamiltonian. Surprisingly it is not too difficult to derive the Legendre transformation $L(x,\dot x)= \dot x^a p_a(x,\dot x) - H_{Zg}(x,p(x,\dot x))$ of $H_{Zg}$ explicitly. The calculations are discussed in appendix~\ref{app:Lagrangian}  and yield
\begin{align}
	\dot x^a p_a 
	&= \frac{g(\dot x, \dot x) + g(\dot x, Z)^2}{\ell g(\dot x, Z)\pm \sqrt{2 \ell^2 g(\dot x, Z)^2 + \ell^2 g(\dot x, \dot x)+ 4}} \nonumber\\
	&-\frac{g(\dot x,Z)}{\ell} \ln\bigg(\frac{1}{2}(\ell (g(\dot x, Z)\pm \sqrt{2 \ell^2 g(\dot x, Z)^2 + \ell^2 g(\dot x, \dot x)+4})\bigg)\label{eq:x(p)}\\
	H_{Zg}(x,p(x,\dot x )) &=\frac{2}{\ell^2} - \frac{g(\dot x, Z)}{\ell}\label{eq:H(x,dotx)} - \frac{4}{\ell^2}\frac{1}{ ( \ell g(\dot x, Z)\pm \sqrt{2 \ell^2 g(\dot x, Z)^2 + \ell^2 g(\dot x, \dot x)+ 4}) } \,.
\end{align}
Even though these expressions are quite involved one can calculate their $\ell \rightarrow 0$ limit and obtain
\begin{align}
	\dot x^a p_a = \frac{1}{2}g(\dot x, \dot x),\quad H_{Zg}(x,p(x,\dot x)) = \frac{1}{4}g(\dot x, \dot x) \Rightarrow L(x,\dot x) = \frac{1}{4}g(\dot x, \dot x)
\end{align}
as expected.

The main qualitative difference between the Hamilton equations in general relativity and the ones for the $\kappa$-deformed Hamiltonian is that in general relativity the $\dot p_a$ equation is only sourced by a term proportional to the Christoffel symbols, while in the $\kappa$-deformed case there are extra source terms. This means that, unlike in general relativity, there exists no coordinate system around every point $q$ of spacetime such that $\dot p_a=0$ at~$q$ (i.e.  it is not possible to define normal coordinates around every point). This nicely demonstrates what we already discussed in Theorem 2 of \cite{Barcaroli:2015xda}, namely that for non-homogeneous Hamiltonians a force-like term appears in the Hamilton equations dragging particles away from auto-parallel motion.

%%%%%%%%%%%%%%%%%%%%%%%%%%%%%%%%%%%%%%%%%%%%%%%%%%%%%%%
\subsection{Observers and redshift}\label{ssec:Redshift}
One prominent feature of physics on curved spacetimes is the gravitational redshift. Following our previous analysis done for homogeneous and isotropic models \cite{Barcaroli:2016yrl}, here we  investigate how the amount of redshift between two observers in a generic curved spacetime is influenced by the $\kappa$-deformation. In order to do so we need a  notion of the frequency $\nu_\sigma(\gamma)$ of a light ray $\gamma$ measured by an observer $\sigma$.

A light ray is a solution $\gamma(\tau) = (x_\gamma(\tau) , p_\gamma(\tau) )$ of the Hamilton equations of motion which satisfies $H(\gamma)=0$. An observer is a curve $\sigma(\lambda) =(x_\sigma(\lambda) , p_\sigma(\lambda) )$ to which a tangent vector is associated via $\dot x_\sigma^a = \bar{\partial}^a H(\sigma)$ and which satisfies the following properties:
\begin{enumerate}
	\item The energy of an observer is real for all masses and spatial momenta, i.e. $H(\sigma)<0$,
	\item It is normalized, i.e. $H(\sigma)=-m_\sigma^2 = constant$,
\end{enumerate}
These conditions are the same conditions observers satisfy in general relativity, which can be realized in the $\ell\to 0$ limit of the theory we are discussing. Note that we do not demand the observer's curve to be a solution of the remaining Hamilton equations, since there exist observers who are not freely falling on spacetime. However the relation between the observer's four momentum $p_\sigma$ and the observer's tangent $\dot x_\sigma$ is given via the first Hamilton equation of motion. So the observer is also subject to the $\kappa$-deformed dynamics, in contrast to other models considered \cite{Jacob:2008bw,Rosati:2015pga,Amelino-Camelia:2016fuh}, in which the observer is formalized just as a low-energetic (classical) worldline. In our case, however, since we are describing deformations to the particles' dynamics in a Schwarzschild-like framework later, the mass of the observer plays a crucial role, being proportional to the influence of the $\kappa$-deformation detected in the observer's reference frame, as we will see explicitly in \ref{ssec:schwarzschildredshift}.

The frequency an observer associates to the light ray is given by
\begin{align}
	\nu_\sigma(\gamma) =  p_{\gamma a} \frac{\dot x_\sigma^a}{m_\sigma} = p_{\gamma a}\frac{\bar{\partial}^a H(\sigma)}{m_\sigma}\,.
\end{align}
Surely this expression only makes sense when the light ray and the observer intersect at a certain point on spacetime.

For the $\kappa$-Poincar\'e Hamiltonian $\dot x$ is displayed in \eqref{eq:dotx} so
\begin{align}
	\nu_\sigma(\gamma) m_\sigma
	&= Z(p_{\gamma})\bigg[-\frac{2}{\ell} \sinh\bigg(\ell Z(p_\sigma)\bigg) + \ell e^{\ell Z(p_\sigma)} (g^{-1}(p_\sigma,p_\sigma) + Z(p_\sigma)^2) \nonumber\\
	&+ 2 e^{\ell Z(p_\sigma)}  Z(p_\sigma)\bigg] +  e^{\ell Z(p_\sigma)} 2 g^{-1}(p_{\sigma }, p_\gamma)
\end{align}
with correct classical limit $\ell \rightarrow 0$
\begin{align}
	\nu_\sigma(\gamma)  =\frac{2}{m_\sigma} g^{-1}(p_{\sigma }, p_\gamma)\,.
\end{align} 
We demanded that $H(\sigma)=- m_\sigma^2$ is constant thus we can use
\begin{align}
	-\frac{4}{\ell^2}\sinh\bigg(\frac{\ell}{2}Z(p_\sigma)\bigg)^2 + e^{\ell Z(p_\sigma)} (g^{-1}(p_\sigma,p_\sigma) + Z(p_\sigma)^2) = -m_\sigma^2
\end{align}
to simplify the frequency to
\begin{align}\label{eq:frequency}
	\nu_\sigma(\gamma) 
	&=\frac{1}{m_\sigma}Z(p_{\gamma})\bigg[\frac{2}{\ell}e^{-\ell Z(p_\sigma)} - \frac{2}{\ell}  - \ell m_\sigma^2 + 2 e^{\ell Z(p_\sigma)}  Z(p_\sigma)\bigg] +  e^{\ell Z(p_\sigma)} \frac{2}{m_\sigma} g^{-1}(p_{\sigma }, p_\gamma)\,.
\end{align}
This last expression can easily be used to calculate the redshift between two different observers $\sigma_1$ and $\sigma_2$ who intersect the light ray at different spacetime positions
\begin{align}
	z+1 = \frac{\nu_{\sigma_1}(\gamma)}{\nu_{\sigma_2}(\gamma)}\,.
\end{align}
In section \ref{ssec:schwarzschildredshift}  we will use this formula to derive the deformation of the gravitational redshift in a $\kappa$-deformation of Schwarzschild geometry.

%%%%%%%%%%%%%%%%%%%%%%%%%%%%%%%%%%%%%%%%%%%%%%%%%%%%%%%
\section{Spherically symmetric $\kappa$-deformed phase space}\label{sec:spsym}
In our previous article \cite{Barcaroli:2015xda} we gave a detailed account of the notion of symmetry in Hamilton geometry. Summarizing, a Hamiltonian $H(x,p)$ is invariant under the action of certain diffeomorphisms $\Phi$ on phase  space if the vector field $X_\Phi$ which induces this diffeomorphism annihilates the Hamiltonian 
\begin{align}
X_\Phi(H) = 0.
\end{align}
Particularly interesting are those diffeomorphisms of phase space which are induced by a diffeomorphism of the spacetime manifold. In this case the symmetry condition becomes 
\begin{align}\label{eq:symmcon}
X^C(H)\equiv(\xi^a\partial_a-p_q\partial_a\xi^q\bar{\partial}^a)H=0\,,
\end{align}
where $X=\xi^a(x)\partial_a$ is the vector field which induces the diffeomorphism of spacetime. The details of the derivation of this symmetry condition can be found in \cite{Barcaroli:2015xda}, while an application  in the context of homogeneous and isotropic geometries is discussed in~\cite{Barcaroli:2016yrl}. In the following we use this construction to define general spherically symmetric Hamiltonians.

%%%%%%%%%%%%%%%%%%%%%%%%%%%%%%%%%%%%%%%%%%%%%%%%%%%%%%%
\subsection{The general case}\label{sec:spsymgen}
In order to study spherically symmetric phase spaces it is most convenient to use spherical coordinates $(t,r,\theta, \phi, p_t, p_r, p_\theta, p_\phi)$. The generators of rotations of spacetime are 
\begin{align}
X_1 &= \sin\phi \,\partial_\theta + \cot\theta \cos\phi \,\partial_\phi \label{eq:KVF2a}\\
X_2 &= -\cos\phi \,\partial_\theta + \cot\theta \sin\phi \,\partial_\phi \label{eq:KVF2b}\\
X_3 &= \partial_\phi.\label{eq:KVF2c}
\end{align}
Their complete lifts are displayed in the appendix \ref{app:Lifts}. Evaluating equation \eqref{eq:symmcon} we find, with the same techniques already used in the homogeneous and isotropic case \cite{Barcaroli:2016yrl}, that the most general spherically symmetric Hamiltonian must take the form
\begin{align}\label{eq:sphHam}
	H(x,p) = H(t, p_t, r, p_r, w(\theta, p_\theta, p_\phi) )\textrm{ with } w^2 = p_\theta^2 + \frac{1}{\sin\theta^2}p_\phi^2\,.
\end{align}
As one could expect, the form of the Hamiltonian is less constrained compared to the homogeneous and isotropic case \cite{Barcaroli:2016yrl}. This freedom translates to the appearance of several free functions in the most general third-order polynomial expansion around the standard metric dispersion relation:
\begin{align}
	H(x,p) 
	&= - A(t,r) p_t^2 + C(t,r) p_t p_r + B(t,r) p_r^2 + R(t,r) w^2 \\
	&+ \ell \left( D(t,r) p_t^3 + E(t,r) p_t^2 p_r + F(t,r) p_t p_r^2 + G(t,r) p_r^3 + J(t,r) p_t w^2 + K(t,r) p_r w^2 \right) + \mathcal{O}(\ell^2)\nonumber\,.
\end{align}

Since we are interested in building a Hamiltonian that reduces to the $\kappa$-Poincar\'e one in the local frame we will have a reduced freedom compared to this general case. In particular, we want to construct a Hamiltonian that, besides having spherical symmetry, can be written in the form~\eqref{eq:kappacov}. The general $\kappa$-deformed Hamiltonian~\eqref{eq:kappacov} is built out of two elements: a spacetime metric term $g^{-1}(p,p)$ and a vector field term $Z(p)$. The mostly considered spherically-symmetric metric term, which contains all spherically symmetric vacuum solutions of the Einstein equations, can be written, after an appropriate choice of coordinates, as\footnote{The most general version of the term would be $g^{-1}(p,p) = - a(t, r) p_t^2 + c(t,r) p_t p_r + b(t, r) p_r^2 + d(t,r) w^2$. In case the gradient of $d(r,t)$ is spacelike or timelike the form we displayed can be achieved, however in case the gradient of $d(r,t)$ is vanishing or a null vector, this may not be possible \cite{Plebanski}.}
\begin{align}
	g^{-1}(p,p) = - a(t, r) p_t^2 + b(t, r) p_r^2 + \frac{1}{r^2} w^2\,.
\end{align}
On the other hand, in order to respect spherical symmetry, the vector field term must take the form
 \begin{align}\label{eq:SphericalZ}
 	Z(p) = c(t,r) p_t + d(t,r) p_r\,,
 \end{align}
subject to the condition $g(Z,Z) = -1$, which yields
\begin{align}\label{eq:vf}
	 - \frac{c(t,r)^2}{a(t,r)} + \frac{d(t,r)^2}{b(t, r)} = -1\,.
\end{align}
 Plugging these objects into the $\kappa$-deformed Hamiltonian \eqref{eq:kappacov} results in the most general spherically symmetric $\kappa$-deformed Hamiltonian:
\begin{align}
	H_{Zg} = -\frac{4}{\ell^2}\sinh\bigg(\frac{\ell}{2}( c p_t + d p_r)\bigg)^2 + e^{\ell( c p_t + dp_r)}((-a+c^2)p_t^2 + 2 cd p_r p_t + (b+d^2)p_r^2 + \frac{1}{r^2} w^2)\,,
\end{align}
where we suppressed the arguments of the functions $a,b,c,d$ for the sake of readability. 

The functions $c$ and $d$, intertwined by \eqref{eq:vf}, identify a family of $\kappa$-deformations of the phase space of a spherically symmetric spacetime. One could hope that some fundamental mechanism derived from a complete theory of quantum gravity would single out one specific correct form of the deformation.

One the other hand, if one restricts to specific spherically-symmetric spacetimes, it is not always the case that there exists such freedom in the definition of the $\kappa$-deformation. For example, including further symmetries like in the homogeneous and isotropic case discussed in \cite{Barcaroli:2016yrl}, the only normalized homogeneous and isotropic vector field evaluated on a $1$-form $P=p_adx^a$ is
\begin{align}
	Z(p) = p_t\,.
\end{align}
Then the unique homogeneous and isotropic $\kappa$-deformed Hamiltonian was found to be
\begin{align}
	H_{qFLRW} = -\frac{4}{\ell^{2}}\sinh\bigg(\frac{\ell}{2} p_t \bigg)^2 +  e^{\ell p_t} a(t)^{-2}\bigg((1 - kr^2)p_r^2 + \frac{1}{r^2} w^2\bigg)\,.
\end{align}
Here no additional degrees of freedom in addition to the scale factor of the $FLRW$ metric, which is determined by the Einstein equations, appear. %Thus it is not necessary to label this Hamiltonian by the vector field $Z$ since $Z$ is unique.
 
In the following we specialize to the $\kappa$-deformation of the most famous spherically symmetric solution of Einstein's equations, the Schwarzschild geometry.
 
%%%%%%%%%%%%%%%%%%%%%%%%%%%%%%%%%%%%%%%%%%%%%%%%%%%%%%%
\subsection{The $\kappa$-deformation of Schwarzschild geometry} \label{ssec:kSchw}
In the Schwarzschild solution of general relativity the functions which determine the spacetime metric are
\begin{align}
	a(t,r) = \frac{1}{1 - \frac{r_s}{r}},\quad b(t,r) = a(t,r)^{-1} = 1 - \frac{r_s}{r}\,,
\end{align}
where $r_{s}$ is the Schwarzschild radius.
Thus the functions $c$ and $d$ appearing in the timelike vector field $Z$ which defines the deformation of the classical phase space, eq. \eqref{eq:SphericalZ}, must satisfy
\begin{align}\label{eq:cdcond}
	- \bigg(1 - \frac{r_s}{r}\bigg)c(t,r)^2 + \frac{d(t,r)^2}{\bigg(1 - \frac{r_s}{r}\bigg)} = -1\,,
\end{align}
according to equation \eqref{eq:vf}. Following the discussion of the previous section we can write down the general spherically symmetric $\kappa$-deformation of the phase space of Schwarzschild spacetime
\begin{align}
	H_{ZSchw}(x,p) = 
	&-\frac{4}{\ell^2}\sinh\bigg(\frac{\ell}{2}( c p_t + d p_r)\bigg)^2 \nonumber\\
	&+ e^{\ell( c p_t + dp_r)}\bigg[\bigg(-\frac{1}{1 - \frac{r_s}{r}}+c^2\bigg)p_t^2 + 2 cd p_r p_t + \bigg( 1 - \frac{r_s}{r}+d^2\bigg)p_r^2 + \frac{1}{r^2} w^2\bigg]\,.
\end{align}
In the rest of this section we omit the subscript ${}_{ZSchw}$ for the sake of readability. As already mentioned we find a family of deformations defined by the function $c$ and $d$ subject to the condition~\eqref{eq:cdcond}. This result demonstrates the importance of our general construction in section \ref{ssec:genkappa}, since without the insight that a vector field parametrizes the possible $\kappa$-Poincar\'e deformations we may not have found this general class of $\kappa$-deformations of Schwarzschild geometry.

%%%%%%%%%%%%%%%%%%%%%%%%%%%%%%%%%%%%%%%%%%%%%%%%%%%%%%%
\subsection{Motion in phase space}\label{ssec:kSchwMotion}
To study observable consequence of the $\kappa$-deformation of Schwarzschild geometry we now discuss the equations of motion for point particles. 

In general relativity the Einstein vacuum equations guarantee that every spherically symmetric solution of the equations is static, also known as Birkhoff's theorem. Since so far we have not developed further the dynamics which the $\kappa$-deformation of a classical spacetime geometry has to satisfy, in the following we assume for simplicity that $c$ and $d$ do not depend on $t$, i.e. that $\partial_t$ induces yet another symmetry of $H$.

Due to the symmetry of the geometry which we are studying there exist several constants of motion, one for each generator of symmetry $X_{I}$, displayed in equations \eqref{eq:KVF2a} to \eqref{eq:KVF2c}, to which we add the generator of time translations $\partial_t$. The constants of motion are found as $X_I(P) = X_I^a(x)p_a$. In fact, it is easy to see that this object is constant along the solutions to the Hamilton equations of motion. One then finds the constants of motion:
\begin{align}
	E = p_t,\ L= p_\phi,\ K_1 = \sin\phi p_\theta + \cot\theta \cos\phi p_\phi, \ K_2 = -\cos\phi p_\theta + \cot\theta \sin\phi p_\phi\,.
\end{align}
We can use these constants to restrict the motion of particles to the equatorial plane, fixing $\theta = \frac{\pi}{2}$ and $p_\theta = 0$. For this case $L=p_\phi=w$. Moreover $H$ itself is another constant of motion representing the dispersion relation
\begin{align}\label{eq:disprel}
	-m^2 =
	&-\frac{4}{\ell^2}\sinh\bigg(\frac{\ell}{2}( c p_t + d p_r)\bigg)^2 \nonumber\\
	&+ e^{\ell( c p_t + dp_r)}\bigg[\bigg(-\frac{1}{1 - \frac{r_s}{r}}+c^2\bigg)p_t^2 + 2 cd p_r p_t + \bigg( 1 - \frac{r_s}{r}+d^2\bigg)p_r^2 + \frac{1}{r^2} w^2\bigg]\,.
\end{align}
Under these conditions the non-trivial Hamilton equations of motion are
\begin{align}
	\dot t = \bar{\partial}^t H,\quad \dot p_r = - \partial_r H,\quad  \dot r = \bar{\partial}^r H,\quad \dot \phi = \bar{\partial}^\phi H\,.
\end{align}

Solving analytically the equations of motion is not possible, so, in order to get a first impression of the sort of  effects caused by  $\kappa$-deformations of Schwarzschild spacetime geometry we choose $c=\frac{1}{\sqrt{|1-\frac{r_s}{r}|}}$ in the region $r>r_s$, i.e. outside the classical horizon, for which equation \eqref{eq:cdcond} implies $d=0$. 
A thorough analysis of the implications of general  $\kappa$-deformations of Schwarzschild geometry, parametrized by the functions $c$ and $d$, will be discussed in an upcoming separate article.

%%%%%%%%%%%%%%%%%%%%%%%%%%%%%%%%%%%%%%%%%%%%%%%%%%%%%%%
\subsection{Observable effects in $d=0$ $\kappa$-deformed Schwarzschild geometry} \label{sub:observable}
Choosing $c=\frac{1}{\sqrt{|1-\frac{r_s}{r}|}}\equiv \frac{1}{\sqrt{A}}$, $r>r_s$ and thus $d=0$,  the $\kappa$-deformed Schwarzschild  Hamiltonian takes the form:
\begin{align}\label{eq:Hd=0}
H(x,p) = 
&-\frac{4}{\ell^2}\sinh\bigg(\frac{\ell}{2}\frac{p_t}{\sqrt{A}}\bigg)^2  +  e^{\frac{ \ell p_t}{\sqrt{A}}}\bigg(  A p_r^2 + \frac{1}{r^2} w^2\bigg)\,.
\end{align}
Using this specific choice of the free functions allows to study some relevant features of the model explicitly. In the following we focus on  the effects of the deformation on the circular orbits around the origin with radius larger than $r_s$, and on the redshift between stationary observers.

%%%%%%%%%%%%%%%%%%%%%%%%%%%%%%%%%%%%%%%%%%%%%%%%%%%%%%%
\subsubsection{Circular particle motion}
The relevant Hamilton equations in the study of circular motion are the ones associated to the radial coordinate and momentum. Moreover, the on-shell condition $H=-m^{2}$ relates the particle's energy $p_{t}$ to the radial and angular momenta:
\begin{align}
 \frac{p_t}{\sqrt{A}} =& -\frac{1}{\ell}\ln\bigg(1 + \frac{\ell^2 m^2}{2} \pm  \ell \sqrt{m^2\bigg(\frac{\ell^2 m^2}{4} + 1\bigg) + \frac{w^2}{r^2} + p_r^2 A}\bigg)\nonumber\\
& \rightarrow  -\frac{1}{\ell}\ln\bigg(1 + \frac{\ell^2 m^2}{2} +  \ell \sqrt{m^2\bigg(\frac{\ell^2 m^2}{4} + 1\bigg) + \frac{w^2}{r^2} + p_r^2 A}\bigg) \label{eq:e^pt}\,,
\end{align}
where the sign was chosen so to have $(\frac{p_{t}}{\sqrt{A}})^2 = - m^2$ for observers with $p_r=w=0$ in the $\ell=0$ limit.

A circular orbit is characterized by constant radial coordinate, $\dot r=0$. Then from the Hamilton equation for $\dot r$ it follows that the radial momentum must be constantly vanishing: 
\begin{align}
0=\dot r = \bar{\partial}^rH =2 A \,p_{r} \,e^{\frac{\ell p_t}{\sqrt{A}}} \Rightarrow p_{r} = 0 \,.
\end{align}
This of course also implies that $\dot p_{r}=0$. Using the  Hamilton equation for the radial momentum:\begin{align}\label{eq:dotpr}
0=\dot p_r =-\partial_r H&=- \frac{1}{\ell}\sinh\bigg( \frac{ \ell p_t}{\sqrt{A}}\bigg)\frac{r_s}{r^2}\frac{p_t}{A^{\frac{3}{2}}} - e^{ \frac{\ell p_t}{\sqrt{A}}}\bigg(\frac{r_s}{r^2}p_r^2 - \frac{2 w^2}{r^3}\bigg) + \frac{\ell}{2}\frac{p_t}{A^{\frac{3}{2}}}\frac{r_s}{r^2}e^{ \frac{\ell p_t}{\sqrt{A}}} \bigg(  A p_r^2 +\frac{w^{2}}{r^2}\bigg)\\
& =- \frac{1}{\ell}\sinh\bigg( \frac{ \ell p_t}{\sqrt{A}}\bigg)\frac{r_s}{r^2}\frac{p_t}{A^{\frac{3}{2}}} + e^{ \frac{\ell p_t}{\sqrt{A}}} \frac{2 w^2}{r^3} + \frac{\ell}{2}\frac{p_t}{A^{\frac{3}{2}}}\frac{r_s}{r^4}w^{2}e^{ \frac{\ell p_t}{\sqrt{A}}} \,, 
\end{align}
where in the second line we used $p_{r}=0$.
Before solving for $r$, we can simplify this expression further by using the mass-shell constraint \eqref{eq:e^pt} to remove the $p_{t}$ dependence:
\begin{align}\label{eq:massiver}
\frac{ \frac{1}{\ell}\ln\bigg(1 + \frac{\ell^2 m^2}{2} +  \ell P_{m}\bigg)}{A} \,\,2r_{s} r  P_{m}\left(1+ \frac{\ell^2 m^2}{2}  +\ell \, P_{m}\right)- 4 w^2=0   \,, 
\end{align}
where we multiplied everything by  $2 r^{3}e^{- \frac{\ell p_t}{\sqrt{A}}}$ and we defined $P_{m}=\sqrt{m^2\bigg(\frac{\ell^2 m^2}{4} + 1\bigg) + \frac{w^2}{r^2} }$.
In the massless limit this becomes:
\begin{align}\label{eq:masslessr}
\frac{ \frac{1}{\ell}\ln\bigg(1 +  \ell \frac{w}{r}\bigg)}{A} \,\,2r_{s} w\left(1+\ell \, \frac{w}{r}\right)- 4 w^2=0   \,, 
\end{align}

In general, the equations \eqref{eq:massiver} and \eqref{eq:masslessr} are not solvable analytically, so we continue our study perturbatively. The above equations read, up to first order in $\ell$:
\begin{align}\label{eq:massiverpert}
\frac{r_{s}}{r-r_{s}}\left(  2m^{2}r^{2}+2w^{2}\left(3-2\frac{r}{r_{s}}\right)+\ell \left( w^{2}+2m^{2}r^{2}\right)\sqrt{m^{2}+\frac{w^{2}}{r^{2}}}  \right)=0  \,, 
\end{align}
for the massive case, and
\begin{align}\label{eq:masslessrpert}
w^{2}\left( -4+2\frac{r_{s}}{r-r_{s}}+\ell w \frac{r_{s}}{r(r-r_{s})} \right)=0   \,
\end{align}
for the massless case.
Solving for $r$ one finds the radius of circular orbits for massive particles:
\begin{align}
r_{m} =  \frac{w^{2}}{m^{2}r_s}\bigg( 1 - \sqrt{1- 3 \left(\frac{r_s m}{w}\right)^{2}}\bigg) + \frac{\ell}{4}w^{2}\,m\sqrt{1 + \left(\frac{w}{m\, r_m^{0}}\right)^{2}} \frac{(4 r_m^{0} - 5 r_s )}{(w^{2} - r_m^{0} r_s m^{2})}\,,
\end{align}
where $r_m^{0} = \underset{\ell \rightarrow 0}{\lim}\ r_m$. In the massless limit this becomes:
\begin{align}\label{eq:innerr}
	r_{m=0} = \frac{3}{2} r_s + \ell \frac{w}{6}\,.
\end{align}
This last results indicates that the photon sphere, which is universal in Schwarzschild geometry, is in fact dependent on the angular momentum of the photons once the Planck-scale deformation is introduced, so that photons with different energy are allowed to orbit a black hole at different altitudes. Such a modification of the geometry of the photon sphere of spherically symmetric black holes would immediately have an influence on further observables like lensing \cite{Virbhadra:1999nm} and the observation of the shadows of black holes \cite{Grenzebach:2015oea}. These subjects go beyond the scope of this article and will be investigated in the future.
 
 %%%%%%%%%%%%%%%%%%%%%%%%%%%%%%%%%%%%%%%%%%%%%%%%%%%%%%%
\subsubsection{Redshift}\label{ssec:schwarzschildredshift}
Our goal here is to compute the change in the energy of a photon as measured by two different observers, $\sigma_{1}$ and $\sigma_{2}$, at rest. The observers are characterized by their spacetime coordinates and momenta:  $\sigma_i = (x_{\sigma_i}, p_{\sigma_i})$, $i=1,2$. Since the observers are at rest only the time component of their four-momentum is nonzero:  $p_{\sigma_{i}}=(p_{\sigma_{i}t},0,0,0)$. 
In this case the mass-shell constraint given by the Hamiltonian reads:
\begin{align}
	H(x_{\sigma_{i}}, p_{\sigma_{i}}) =  -\frac{4}{\ell^2}\sinh\bigg(\frac{\ell}{2}\frac{p_{\sigma_{i}t}}{\sqrt{A(x_{\sigma_{i}})}}\bigg)^2 =-m^{2}_{\sigma_{i}}\,.
\end{align}
This constraint implies that the four-momentum of the observers is related to their position and mass via $p_{\sigma_i t} = \sqrt{A(x_{\sigma_i})} Q_{\sigma_i}$, with $Q_{\sigma_i}\equiv-\frac{1}{\ell} \ln\left( 1+\frac{\ell^{2}m_{\sigma_{i}}^{2}}{2}+\ell m_{\sigma_{i}}\sqrt{1+\frac{\ell^{2}m^{2}_{\sigma_{i}}}{4}} \right)$  being a constant.

Having defined the observers, we can use equation \eqref{eq:frequency} to obtain the frequencies that each of them associates to the photon:
\begin{align}
	\nu_{\sigma_i}(\gamma) 
	&= \frac{1}{m_{\sigma_{i}}}\frac{p_{\gamma t}|_{\sigma_i}}{\sqrt{A(x_{\sigma_i})}}\bigg[\frac{2}{\ell}e^{-\ell Q_{\sigma_i}} - \frac{2}{\ell}  - \ell m_{\sigma_i}^2 \bigg] \nonumber\\
	&= \frac{p_{\gamma t}|_{\sigma_i}}{\sqrt{A(x_{\sigma_i})}}\left[2 \sqrt{1+\frac{\ell^{2}m^{2}_{\sigma_{i}}}{4}}  \right] 
\end{align}
The time component of the momentum of the photon at the position of the observer $\sigma_{i}$ is given by $p_{\gamma t}|_{\sigma_i}$. Since the light trajectory~$\gamma$ is a solution of the Hamilton equations of motion, $p_{\gamma t}$ is constant along~$\gamma$. In particular, $p_{\gamma t}$ has the same value at the intersection point with $\sigma_1$ and at the intersection point with $\sigma_2$, so $p_{\gamma t}|_{\sigma_1}=p_{\gamma t}|_{\sigma_2}=p_{\gamma t}$.
The redshift of the photon between the two observers is thus given by
\begin{align}\label{eq:redshift}
	z+1 &= \frac{\nu_{\sigma_1}(\gamma) }{\nu_{\sigma_2}(\gamma) } = \frac{\sqrt{A_2}}{\sqrt{A_1}} \,\frac{\sqrt{1+\frac{\ell^{2}m^{2}_{\sigma_{1}}}{4}}
}{\sqrt{1+\frac{\ell^{2}m^{2}_{\sigma_{2}}}{4}} 
} = \sqrt{\frac { 1 - \frac{r_s}{r_2} } { 1 - \frac{r_s}{r_1} } }\, \frac{\sqrt{1+\frac{\ell^{2}m^{2}_{\sigma_{1}}}{4}} 
}{\sqrt{1+\frac{\ell^{2}m^{2}_{\sigma_{2}}}{4}}   
} \\
&\simeq \sqrt{\frac { 1 - \frac{r_s}{r_2} } { 1 - \frac{r_s}{r_1} } } \left(1+\frac{\ell^{2}}{8}(m_{\sigma_1}-m_{\sigma_2})(m_{\sigma_1}+m_{\sigma_2})\right)\,,
\end{align}
where in the last step we only kept the lowest order $\ell$-correction.
Thus for  two static observers the redshift of a photon is identical to the one in Schwarzschild geometry to all orders in $\ell$, if the observers have the same mass. Otherwise, if the observers have different masses, then they  measure a redshift which departs from the standard result to second order in $\ell$. This influence of the mass of the observers on the redshift is due to the fact that we assumed that the observers are also subject to the $\kappa$-deformed dynamics. If one were to assume that observers follow the dynamics of the general relativistic Hamiltonian \eqref{eq:Hg}, or that the observers have negligible masses, then there would be again no additional effect compared to the usual redshift in Schwarzschild geometry.

Surely the results of this section highly depend on the specific choice of observers and of the vector field $Z$ (remember that the possible deformations of Schwarzschild geometries encoded by the vector field $Z$ depend on two free functions of spacetime coordinates, which we fixed at the beginning of this subsection \ref{sub:observable} in order to have a workable example). In general we would expect that the Planck-scale deformation would alter the gravitational redshift of photons in spherical symmetry also for equal-mass observers, as it is the case in the homogeneous and isotropic cosmological situation discussed in~\cite{Barcaroli:2016yrl}.

%%%%%%%%%%%%%%%%%%%%%%%%%%%%%%%%%%%%%%%%%%%%%%%%%%%%%%%
\section{Discussion}
We used the insights we gained in the local implementation of the $\kappa$-Poincar\'e dispersion relation on homogeneous and isotropic spacetimes \cite{Barcaroli:2016yrl} to extend our findings to general curved spacetimes. The key result of our work is the construction of a phase space in which locally one can identify a spacetime with $\kappa$-Lorentz  symmetry, i.e. $\kappa$-Poincar\'e symmetries excluding translations. The implementation of this local symmetry via the level sets of a Hamilton function on the point particle phase space causes the geometry of spacetime and the geometry of momentum space to be intertwined into a geometry of phase space. 

In equation \eqref{eq:kappacov} we presented the locally $\kappa$-Poincar\'e Hamilton function which deserves its name by the fact that at every point on spacetime there exists a local basis of the cotangent spaces of the spacetime manifold such that the level sets of the Hamilton function assume the form of the $\kappa$-Poincar\'e dispersion relation. This is the direct generalization of local Lorentz invariance of the geometry of spacetime to local $\kappa$-Lorentz invariance. The explicit construction of the $\kappa$-Poincar\'e Hamilton function will allow us to study the mathematical differential geometric structure of the phase space geometry in the future. In particular, the local frame bundle properties of spacetime are of interest since equivalent frames are no longer identified with linear transformations like Lorentz transformations but with the partly non-linear $\kappa$-Lorentz transformations, the $\kappa$-Poincar\'e boosts and rotations.

Having established the notion of a general $\kappa$-deformed phase space we studied the motion of test particles on such a background. The modification of the geodesic equation was presented in equation \eqref{eq:deformedgeod}. As already stated when we introduced Hamiltonian geometry in \cite{Barcaroli:2015xda}, there appears a force-like term in the equations of motion which can not be absorbed into the geometry of spacetime. Thus there exists no local coordinate system such that the equations of motion locally reduce to $\ddot x + \mathcal{O}(x^2) = 0$ as they do in normal coordinates in the undeformed spacetime geometry. Also generalizations of normal coordinates, as they were discussed in the context of Finsler geometry in \cite{Pfeifer:2014eva} and \cite{Minguzzi:2016gct}, do not exist. To complete the discussion on particle motion on the $\kappa$-deformed phase space geometry we derived the Lagrangian formulation of point particle motion. This can be used as starting point for the derivation of a Finslerian version of the locally $\kappa$-deformed spacetime geometry in the future, as it was done for particular $\kappa$-deformed geometries in \cite{Amelino-Camelia:2014rga,Lobo:2016xzq,Letizia:2016lew}.

In the second half of this article we derived the most general form of the locally $\kappa$-Poincar\'e Hamilton function compatible with spherical symmetry. We obtained a Hamilton function defined in terms of four free functions of the time and radial coordinate, two of which are fixed by the specific spacetime geometry on which the deformation is based. The presence of the other two free functions is due to the fact that the timelike vector field which is necessary to define the Hamilton function is not fixed by the available symmetry constraints. This is to be contrasted with the homogeneous and isotropic case \cite{Barcaroli:2016yrl}, where the symmetry constraints were sufficient to fully determine the form of the deformation. 

We studied observable predictions of the model in the special case of deformations of the Schwarzschild geometry, where the vector field defining the deformation was chosen as the  tangent of the standard observer at rest in Schwarzschild geometry.  In an upcoming article we will investigate the influence of the choice of this vector field on observables in more detail. The freedom in the choice of the vector field defining the deformed Hamiltonian may be related to the deformed boosts which underly the $\kappa$-deformed spacetime geometry, in the sense that the deformed boost may map one choice of $Z$ to another. This will be matter of investigation in future work.  
For our choice of $\kappa$-deformed Schwarzschild geometry we studied two possibly observable features: the radius of photon orbits around the spherical symmetric black hole (known as photon sphere in the standard case) and the gravitational redshift between two observers at rest with respect to each other and with respect to the black hole horizon. For the first observable we found that the photon sphere, which is universal for all photons in Schwarzschild geometry, becomes momentum dependent. In particular, photons with a different angular momentum have circular orbits at different altitudes \eqref{eq:innerr}. For the redshift we found that corrections to the standard Schwarzschild  case emerge only at the second order in the deformation parameter, \eqref{eq:redshift}. Moreover, these corrections  are proportional to the difference of the masses of the observers measuring the frequency of the photon and they only exist if one assumes that the observers enjoy the same deformed symmetries as the photon itself. 

In an upcoming work we will study the spherically symmetric $\kappa$-Poincar\'e deformed spacetime geometry in further detail to derive observable implications in solar system and black hole observations, like perihelion shifts, light deflections, the horizon and the singularity. Further interesting studies which are now in reach are locally $\kappa$-deformed spacetime geometries with any desired symmetry, like axial symmetry, as generalization of the spherically symmetric case.

Besides these phenomenological studies, one can further develop our method to locally implement more general dispersion relations on curved spacetime, generalizing the $\kappa$-Poincar\'e case that was studied here. The procedure to be applied would be to identify four basis vector fields $\{Z_i\}_{i=0}^3$ on spacetime which represent, when applied to a four momentum $Z_i(p)$, the different Cartesian momentum components  $p_i = Z_i(p)$. This sort of generalization would be particularly interesting since it would allow to compare predictions concerning black hole physics obtained in the framework of Hamilton geometry to the ones obtained using rainbow gravity as a formalization of Planck-scale effects \cite{Ling:2005bp, Ali:2014xqa, Leiva:2008fd, Gim:2014ira}.

%------------------------------------------------------------------------------%
\begin{acknowledgments}
	CP gratefully thanks the Center of Applied Space Technology and Microgravity (ZARM) at the University of Bremen for their kind hospitality and acknowledges partial support of the European Regional Development Fund through the Center of Excellence TK133 ``The Dark Side of the Universe''. GG acknowledges support from the John Templeton Foundation. LKB acknowledges the support by a Ph.D. grant of the German Research Foundation within its Research Training Group 1620 \emph{Models of Gravity}. NL aknowledges partial support from the 000008 15 RS {\it Avvio alla ricerca} 2015 fellowship (by the Italian ministry of university and research).
\end{acknowledgments}

%%%%%%%%%%%%%%%%%%%%%%%%%%%%%%%%%%%%%%%%%%%%%%%%%%%%%%%
\newpage
\appendix
\newpage

%%%%%%%%%%%%%%%%%%%%%%%%%%%%%%%%%%%%%%%%%%%%%%%%%%%%%%%
\section{Theories of electrodynamics leading to $\kappa$-Poincar\'e light propagation}\label{app:kappaelectro}
To demonstrate that the $\kappa$-deformed Hamiltonian we constructed in equation \eqref{eq:kappacov} can be obtained as the geometric optics limit of a theory of electrodynamics we summarize here the arguments leading to such a theory.

Theories of electrodynamics are rooted in field equations which yield charge and magnetic flux conservation. Following the axiomatic approach to electrodynamics presented in \cite{Hehl} a most general way to formulate such a theory requires two $2$-form fields, the electric field strength $F$ and the magnetic excitation $H$, and a closed current $3$-form $J$ subject to the equations
\begin{align}\label{eq:Maxwell}
	d F = 0, \quad dH=J.
\end{align}
In four dimensions these are eight equations which shall determine the components of the fields $F$ and $H$, which are twelve in total. Thus this set of equations is not sufficient to yield a predictive theory of electrodynamics. In addition a so called constitutive relation $\#$ is necessary to define a predictive theory of electrodynamics which defines a functional dependence of the excitation on the field strength
\begin{align}
	H = \#(F)\,.
\end{align}
Combining these equations on a contractable spacetime one obtains $F=dA$, which makes the theory a theory with a $1$-form potential $A$ as fundamental field and gauge invariance. The remaining field equation, the second in \eqref{eq:Maxwell}, becomes $d\#dA = J$ as dynamical equation for the potential. All theories of electrodynamics constructed according to this scheme are gauge invariant by construction.

The most famous examples of theories of electrodynamics are local and linear, i.e. $H$ is a linear function of $F$. In Maxwell vacuum electrodynamics on curved spacetime the constitutive relation is given by the Hodge star operator of the metric $H = \star F$, or in components $H_{ab} = \frac{1}{2}\epsilon_{abcd}g^{ce}g^{df}F_{ef}$, while for example electrodynamics in media is described by a general local and linear constitutive relation $H_{ab} = \frac{1}{4}\epsilon_{abcd}\chi^{cdef}F_{ef}$. Here $\epsilon_{abcd}$ is the Levi-Civita symbol and $\chi^{abcd}$ the so called constitutive density, where we omit explicit displaying density factors like determinants of the metric, for the sake of a compact presentation of the arguments. Details on this approach to electrodynamics can be found for example in \cite{Hehl,Hehl:2002hr,Pfeifer:2016har} and further references therein.

To obtain a theory of electrodynamics which implies propagation of light (resp. propagation of singularities in the language of partial differential equations \cite{Hoermander1,Hoermander2,Hoermander3,dencker:1982}) governed by the $\kappa$-deformed Hamiltonian we consider the following class of linear higher derivative constitutive relations
\begin{align}\label{eq:kappaconlaw}
	H_{ab} = \frac{1}{2}\epsilon_{abcd} G_{(Q,S)}^{ec}(x,\partial)G_{(Q,S)}^{df}(x,\partial)F_{df}
\end{align}
with
\begin{align}\label{eq:gpartial}
	G_{(Q,S)}^{ab}(x,\partial) = \frac{4}{\ell^2}\sinh\bigg(i\frac{\ell}{2}Z(\partial)\bigg)^2 \frac{Q^{ab}}{Q(\partial,\partial)}- e^{-\ell i Z(\partial)} (g^{-1}(\partial,\partial) + Z(\partial)^2)\frac{S^{ab}}{S(\partial,\partial)}\,,
\end{align}
where $Z(\partial) = Z^a(x)\partial_a$, $g^{-1}(\partial,\partial) = g^{-1ab}(x)\partial_a\partial_b$, $Q(\partial,\partial) = Q^{ab}(x,\partial)\partial_a \partial_b$ and $S(\partial,\partial) = S^{ab}(x,\partial)\partial_a\partial_b$. The operators $Q$ and $S$ parametrize different constitutive relations and thus different theories of electrodynamics. Simple choices, not involving further derivatives, may be
\begin{align}
	Q^{ab} = g^{ab} = S^{ab},\quad \text{or}\quad Q^{ab} = g^{ab},\ S^{ab} = g^{ab} + Z^a Z^b\,.
\end{align}
As a remark recall that using constitutive laws which involve derivative operators is something known in the literature. The most famous example of such a higher derivative theory of electrodynamics may be Bopp-Podolski electrodynamics \cite{Bopp,Podoslky}, which is studied as a candidate theory of electrodynamics which yields a finite self force of charged particles \cite{Gratus:2015bea}.

We will now demonstrate that all theories of electrodynamics which are constructed from a constitutive law of the form \eqref{eq:kappaconlaw} yield wave propagation governed by the Hamiltonian \eqref{eq:kappacov}.

It is well known that for local and linear constitutive laws the wave propagation is governed by the Fresnel polynomial, first derived in \cite{Hehl:2002hr},
\begin{align}\label{eq:Fresnel}
  \mathcal{G}(x,p) = \frac{1}{4!} \epsilon_{c_1 a_1 a_2 a_3} \epsilon_{d_3 b_1 b_2 b_3} \chi^{a_1 c_1 b_1 d_1}(x) \chi^{a_2 c_2 b_2 d_2}(x) \chi^{a_3 c_3 b_3 d_3}(x) p_{d_1} p_{c_2} p_{d_2} p_{c_3}\,.
\end{align}
It serves as Hamiltonian which determines the motion of light along those solutions of Hamiltons equations of motion which satisfy $\mathcal{G}(x,p)=0$. Technically speaking it is the principal polynomial of the dynamical equation $d\#(dA) = J$, obtained from its Fourier space representation. Or, in other words, the highest derivative term in the equation where the partial derivatives are exchanged with~$- i p$. Since only this highest order derivative term is relevant for the geometric optics limit of the theory we do not need to worry about using covariant or partial derivatives when defining \eqref{eq:kappaconlaw} in terms of \eqref{eq:gpartial}. Terms involving connection coefficient, which covariantize the field equations, are of lower order derivatives acting on the dynamical field and thus do not contribute.

In Maxwell electrodynamics with $\chi^{abcd} \sim g^{a[c}g^{d]b}$ one obtains $\mathcal{G}(x,p) = (g^{ab}p_ap_b)^2$ while for example in an uniaxial crystal with $\chi^{abcd} \sim g^{a[c}g^{d]b} + U^{[a}X^{b]}U^{[c}X^{d]}$, where $X$ denotes the crystal axis and $U$ the rest frame of the crystal, one obtains birefringent light propagation from the bi-metric Fresnel polynomial $\mathcal{G}(x,p) = g^{ab}p_ap_b(g^{cd}- (g_{ij}X^iX^j) U^cU^d + X^c X^d)p_cp_d$.

Algebraically we are dealing with the same situation as in the general local and linear case, except that our constitutive relation is a partial differential operator. Following the usual derivation of the Fresnel polynomial, again see \cite{Hehl,Hehl:2002hr,Pfeifer:2016har} for details, we find the same algebraic expression for the principal symbol except that our constitutive density now depends on momenta, by the interchange of $\partial \rightarrow -i p$, when going from configuration to Fourier space
\begin{align}\label{eq:Fresnel2}
\mathcal{G}(p) = \frac{1}{4} \epsilon_{c_1 a_1 a_2 a_3} \epsilon_{d_3 b_1 b_2 b_3} \chi^{a_1 c_1 b_1 d_1}(-ip) \chi^{a_2 c_2 b_2 d_2}(-ip) \chi^{a_3 c_3 b_3 d_3}(-ip) p_{d_1} p_{c_2} p_{d_2} p_{c_3}\,.
\end{align}
Since the specific constitutive relation is constructed from an operator which has the same index structure as the Hodge star of a metric it is simple to calculate the Fresnel polynomial \eqref{eq:Fresnel2} for the constitutive relation \eqref{eq:kappaconlaw} and we find
\begin{align}
	\mathcal{G}(p) = (G^{ab}(x,-ip)p_a p_b)^2 = (H_{Zg}(x,p))^2\,,
\end{align}
for any choice of $Q$ and $S$. An explicit calculation can be found in \cite{Pfeifer:2016har} for the standard Maxwell case.

Thus there exists a huge class of higher derivative gauge invariant theories of electrodynamics, parametrized by $Q$ and $S$, whose geometric optics limit is governed by the locally $\kappa$-deformed Hamiltonian we constructed in this article and is thus invariant under local $\kappa$-Poincar\'e transformations.

As final remark we like to point put that the symmetries of the geometric optic limit and the full field theory may very well differ. In the case of the uniaxial crystal, the field equations are defined in terms of a metric and two vector fields. In the geometric optics limit these building blocks combine to a bi-metric Fresnel polynomial, thus the geometric optics posses all the symmetries these metrics share. The full field theory however can not be formulated in terms of the two metrics alone and hence may posses different symmetries.
Thus whether the theory of electrodynamics leading to $\kappa$-deformed geometric optics must be locally $\kappa$-Poincar\'e invariant itself is an open issue. It may very well be that local $\kappa$-Poincar\'e invariance is only a geometric optics feature and not one of the full field theory. This would in particular depend on the full quantum gravity theory whose semiclassical limit can be described in terms of the local $\kappa$-Poincar\'e symmetries.

%%%%%%%%%%%%%%%%%%%%%%%%%%%%%%%%%%%%%%%%%%%%%%%%%%%%%%%
\section{The $\kappa$-Poincar\'e Lagrangian}\label{app:Lagrangian}
In section \ref{ssec:kappamotion} we discussed the Hamilton equations of motion of the general $\kappa$-Poincar\'e Hamiltonian. Here we demonstrate how the corresponding Lagrangian can be obtained from which one can derive the second oder Euler-Lagrange equations. The Legendre transformation form the Hamiltonian to the Lagrangian involves the terms
\begin{align}
	L(x,\dot x) = \dot x(p(x,\dot x)) - H(x, p(x,\dot x))
\end{align}
which we will derive now. 

In \eqref{eq:dotx} we already found
\begin{align}
\dot x^a &= \bar{\partial}^a H\\
&= Z^a \bigg[-\frac{2}{\ell} \sinh\bigg(\ell Z(p)\bigg) + \ell e^{\ell Z(p)} (g^{-1}(p,p) + Z(p)^2) + 2 e^{\ell Z(p)}  Z(p)\bigg] +  e^{\ell Z(p)} 2 g^{ab}p_b\,. \nonumber
\end{align}
Contracting this equation with $Z$ yields
\begin{align}\label{eq:g(dot x,z)}
g(\dot x, Z) = \frac{2}{\ell} \sinh\bigg(\ell Z(p)\bigg) - \ell e^{\ell Z(p)} (g^{-1}(p,p) + Z(p)^2) 
\end{align}
which allows us to write
\begin{align}\label{eq:dotxp}
\dot x^a &= Z^a \bigg[-g(\dot x, Z) + 2 e^{\ell Z(p)}  Z(p) \bigg] + e^{\ell Z(p)} 2 g^{ab}p_b \,,
\end{align}
and
\begin{align}
\dot x^a p_a = - Z(p) g(\dot x, Z) + 2 e^{\ell Z(p)} (Z(p)^2+g^{-1}(p,p))\,.
\end{align}
as well as
\begin{align}
g(\dot x, \dot x) &= - g(\dot x, Z)^2 + 2 e^{\ell Z(p)}(Z(p)g(\dot x,Z)+ \dot x(p))\\
&=- g(\dot x, Z)^2 + 4 e^{ 2\ell Z(p)} (Z(p)^2+g^{-1}(p,p))\\
&=- g(\dot x, Z)^2 + 4 e^{ \ell Z(p)}\bigg(\frac{2}{\ell^2}\sinh(\ell Z(p))-\frac{g(\dot x, Z)}{\ell}\bigg)\\
&= - g(\dot x, Z)^2  -\frac{4}{\ell} e^{\ell Z(p)}g(\dot x, Z)+\frac{4}{\ell^2}(e^{2\ell Z(p)}-1)
\end{align}
The last equation can be reformulated as quadratic equation for $e^{\ell Z(p)}$
\begin{align}
0=e^{2\ell Z(p)} - \ell e^{\ell Z(p)}g(\dot x, Z) - \frac{\ell^2}{4}(g(\dot x, \dot x) + g(\dot x, Z)^2) -1\,.
\end{align}
with solution
\begin{align}
e^{\ell Z(p)} 
&= \frac{\ell}{2}g(\dot x, Z)\pm \sqrt{\frac{\ell^2}{2}g(\dot x, Z)^2+\frac{\ell^2}{4}g(\dot x,\dot x)+1}\\
&=\frac{1}{2} \big( \ell g(\dot x, Z)\pm \sqrt{2 \ell^2 g(\dot x, Z)^2+\ell^2 g(\dot x,\dot x)+4} \big)\label{eq:e^Z}\\
Z(p) &= \frac{1}{\ell}\ln\bigg(\frac{1}{2} \big( \ell g(\dot x, Z)\pm \sqrt{2 \ell^2 g(\dot x, Z)^2+\ell^2 g(\dot x,\dot x)+4} \big)\bigg)\,.
\end{align}
Finally we can use the terms we found to solve \eqref{eq:dotxp} for $p(x,\dot x)$
\begin{align}
p_a (x,\dot x)
&= \frac{1}{2}g_{ab}\dot x^b e^{-\ell Z(p)} -\frac{1}{2}g_{ab} Z^b \bigg[- e^{-\ell Z(p)}  g(\dot x, Z) + 2 Z(p)\bigg]\\
&= \frac{1}{2}e^{-\ell Z(p)}\bigg(g_{ab}\dot x^b  + g_{ab} Z^b g(\dot x, Z) \bigg)  - g_{ab} Z^b Z(p)\\
&=\frac{g_{ab}\dot x^b  + g_{ab} Z^b g(\dot x, Z) }{\ell g(\dot x, Z)\pm \sqrt{2 \ell g(\dot x, Z)^2+\ell g(\dot x,\dot x)+4}} \\
&- \frac{g_{ab}Z^b}{\ell}\ln\bigg(\frac{1}{2} \big( \ell g(\dot x, Z)\pm \sqrt{2 \ell g(\dot x, Z)^2+\ell g(\dot x,\dot x)+4} \big)\bigg)\,.
\end{align}
Contracting this expression with $\dot x^a$ yields the desired equation \eqref{eq:x(p)}. Equation \eqref{eq:H(x,dotx)} is obtained by solving \eqref{eq:g(dot x,z)} for
\begin{align}
	e^{\ell Z(p)} (g^{-1}(p,p) + Z(p)^2)  = \frac{2}{\ell^2} \sinh\bigg(\ell Z(p)\bigg)  - \frac{g(\dot x, Z)}{\ell},
\end{align}
plugging this result into the Hamiltonian \eqref{eq:kappacov} and inserting \eqref{eq:e^Z} afterwards.
	
%%%%%%%%%%%%%%%%%%%%%%%%%%%%%%%%%%%%%%%%%%%%%%%%%%%%%%%
\section{The lifts of the symmetry generating vector fields to phase space}\label{app:Lifts}
In section \ref{sec:spsymgen} we used the lifts of the vector fields which generate spherical symmetry on spacetime to derive the most general spherically symmetric Hamilton function on phase space. These lifts 
\begin{align}
	X^C_I = \xi^a\partial_a-p_q\partial_a\xi^q\bar{\partial}^a
\end{align}
of the vector fields $X_I = \xi^a_I(x)\partial_a, I=1,2,3$ (see equations \eqref{eq:KVF2a} to \eqref{eq:KVF2c}) are given by
\begin{align}
X_1^C &= \sin\phi \partial_\theta + \cot\theta \cos\phi \partial_\phi\nonumber\\
&+ \frac{\cos\phi}{\sin\theta^2}p_\phi \bar{\partial}^\theta - \bigg(\cos\phi p_\theta - \cot\theta \sin\phi p_\phi\bigg)\bar{\partial}^\phi\,,
\end{align}
\begin{align}
X_2^C &= -\cos\phi \partial_\theta + \cot\theta \sin\phi \partial_\phi\nonumber\\
&+ \frac{\sin\phi}{\sin\theta^2}p_\phi \bar{\partial}^\theta - \bigg(\sin\phi p_\theta + \cot\theta \cos\phi p_\phi\bigg)\bar{\partial}^\phi\,,
\end{align}
\begin{align}
X_3^C &= \partial_\phi\,.
\end{align}
It can be easily checked by direct calculation that the Hamiltonian \eqref{eq:sphHam} satisfies $X^C_I(H)=0$ for all $I=1,2,3$.

%%%%%%%%%%%%%%%%%%%%%%%%%%%%%%%%%%%%%%%%%%%%%%%%%%%%%%%
\bibliographystyle{utphys}
\bibliography{GC}

\end{document}